\documentclass{article}

\usepackage[nonatbib,final]{neurips_2018}

\usepackage[utf8]{inputenc} 
\usepackage[T1]{fontenc}    
\usepackage[colorlinks=true,linkcolor=blue,citecolor=blue]{hyperref}       
\usepackage{url}            
\usepackage{booktabs}       
\usepackage{amsfonts}       
\usepackage{nicefrac}       
\usepackage{microtype}      
\usepackage{graphicx}
\usepackage{helvet}
\usepackage{courier}
\usepackage{amsmath}
\usepackage{amssymb}
\usepackage{tabularx}
\usepackage{varwidth}
\usepackage{xcolor}
\usepackage{graphicx}
\usepackage{bbm}
\usepackage{bm}
\usepackage{colortbl}
\usepackage[clock]{ifsym}
\usepackage{enumitem}
\usepackage{multicol} 
\usepackage[utf8]{inputenc}
\usepackage{booktabs}  
\usepackage{algorithm}
\usepackage{algorithmic}
\usepackage{wrapfig}
\usepackage{amsthm}
\usepackage{multirow}
\usepackage{subfig}
\usepackage{setspace}
\usepackage{pifont}
\usepackage[export]{adjustbox}

\newcolumntype{C}[1]{>{\centering\arraybackslash}m{#1}}
\newcolumntype{R}[1]{>{\raggedright\arraybackslash}m{#1}}

\usepackage[font=small,skip=10pt]{caption}
\usepackage{setspace}
\DeclareCaptionFormat{myformat}{#1#2#3\hrulefill}
\newcommand{\Tr}{\text{Tr }}
\newcommand{\Var}{{\textrm{Var}}}
\newcommand{\RMMD}{{\textrm{RMMD}}}
\newcommand{\MMD}{{\textrm{MMD}}}
\newcommand{\HSIC}{{\textrm{HSIC}}}
\newcommand{\RHSIC}{{\textrm{RHSIC}}}

\title{Application of Kernel Hypothesis Testing on Set-valued Data}

\author{
  Alexis Bellot$^{1,2}$\hspace{0.5cm} Mihaela van der Schaar$^{1,2,3}$\\
  $^{1}$University of Cambridge, $^{2}$The Alan Turing Institute, $^{3}$University of California Los Angeles\\
  \texttt{[abellot,mschaar]@turing.ac.uk} \\
}

\begin{document}

\maketitle

\begin{abstract}
We present a general framework for hypothesis testing on distributions of \textit{sets} of individual examples. Sets may represent many common data sources such as groups of observations in time series, collections of words in text or a batch of images of a given phenomenon. This observation pattern, however, differs from the common assumptions required for hypothesis testing: each set differs in size, may have differing levels of noise, and also may incorporate nuisance variability, irrelevant for the analysis of the phenomenon of interest; all features that bias test decisions if not accounted for. In this paper, we propose to interpret sets as independent samples from a collection of latent probability distributions, and introduce kernel two-sample and independence tests in this latent space of distributions. We prove the consistency of these tests and observe them to outperform in a wide range of synthetic experiments. Finally, we showcase their use in practice with experiments on healthcare and climate data, where previously heuristics were needed for feature extraction and testing.
\end{abstract}

\section{Introduction}
Hypothesis tests are used to answer questions about a specific dependency structure in data (e.g. independence between variables, equality of distributions between samples etc). They are used in applications across the sciences where they serve as an essential tool to summarize and quantify the evidence for structure in the distribution of data \cite{lehmann2006testing}. In consequence, a growing body of work is constantly revisiting established modelling assumptions to allow for consistent testing in increasingly heterogeneous data sources. Examples include non-parametric tests formulated as distances in Hilbert space \cite{gretton2008kernel,gretton2009fast, gretton2012kernel,zhang2018large,bellotkernel}, tests based on neural network representations \cite{liu2020learning,lopez2016revisiting,bellot2019conditional} and others that have significantly advanced the reach of hypothesis tests towards high-dimensional data of unknown distribution. 

Almost universally however, non-parametric tests require a \textit{fixed} presentation of data (e.g. each instance living in $\mathbb R^d$) and do not account for \textit{non-homogeneous noise} patterns across examples (e.g. such as found in medical data, each patient or instance having different level of variation). Many problems do exhibit these properties, including time series (e.g. multiple observations over time for each individual) and bagged data (e.g. multiple images of the same phenomenon) in domains such as medicine and climate science. 

\begin{figure*}[t]
\captionsetup{format=myformat}
\centering
\includegraphics[width=0.9\textwidth]{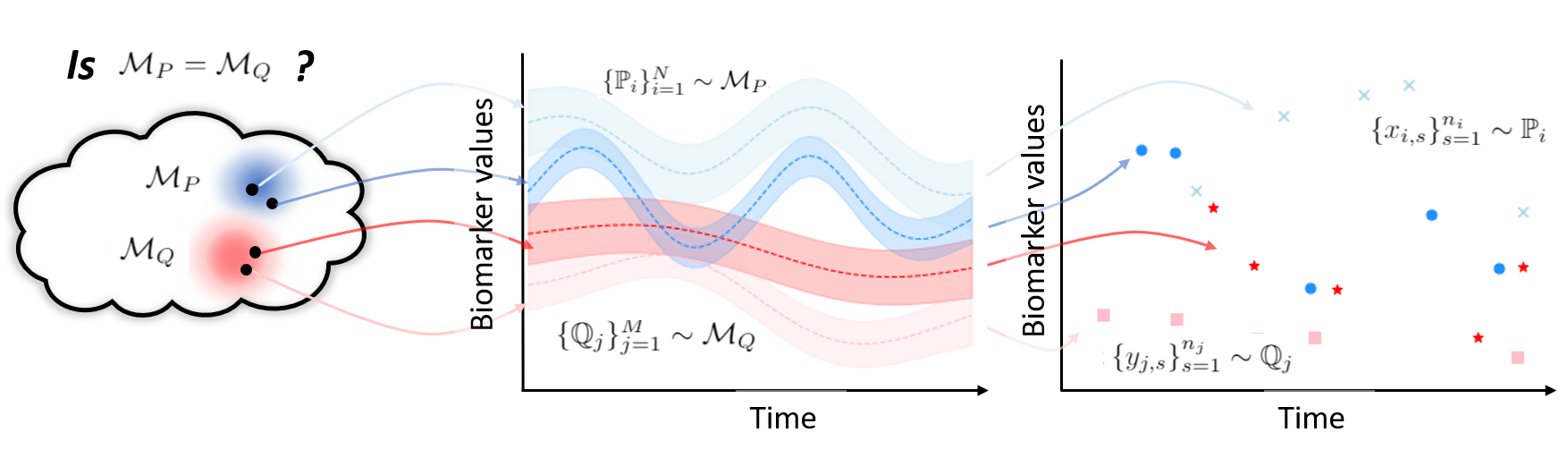}
\caption{We consider an example from electronic health records to illustrate the proposed approach. \textbf{Right panel:} we observe irregular, uncertain biomarker measurements over time in two groups of patients (treated and control) colored with different shades of red and blue, the question being whether these populations have the same trajectory in distribution. \textbf{Middle panel}: we encode the uncertainty in each patient trajectory by a probability distributions on the space of observations. \textbf{Left panel:} The two-sample problem is to test for equality in distribution on the space of patient-specific distributions, rather than actual observations. This two-level hierarchy allows for noisy inputs and irregular input sizes. A description of the notation and more details can be found in Section \ref{two_sample_problem}.}
\label{overview}
\end{figure*}

Intriguingly, there exists an appropriate representation of data that naturally encodes a more flexible observation pattern, namely each example represented as a \textit{set} of observations (i.e. an unordered collection of multivariate observations), each set of potentially irregular length and sampled from potentially different distributions. In particular, sets do not presuppose a fixed representation of data (sets may be of different length) and each set may be associated with a unique distribution that encodes its particular variation pattern (potentially different from other sets). Testing on sets implicitly shifts the question of interest from a hypothesis on groups of actual observations to an hypothesis on groups of latent distributions assumed to represent each observed example or set. See Figure \ref{overview} for an illustration of this interpretation for the two sample problem. This set-up is common in regression problems where one seeks to learn a mapping from distributions to associated labels \cite{szabo2015two,szabo2016learning}, but is unexplored in hypothesis testing. 

The goal of this paper is to introduce kernel two-sample and kernel independence tests defined on \textit{set}-valued examples. 

We will show that tests defined in this space appropriately encode individual-level heterogeneity, are much more flexible, do not require heuristic pre-processing of data, and are found to be more powerful than alternatives. We propose an approach applicable to any kernel-based test that includes, in addition to two-sample and independence tests described here, conditional independence tests and three-variable interaction tests. 

The technical challenge to achieve consistency of test decisions is that latent distributions on which tests are defined are not available (and instead are approximated with each available set of observation). This introduces an additional layer of uncertainty that must be bounded to derive well-defined asymptotic distributions for the proposed test statistics. For this reason, we put emphasis also on the quality of finite-dimensional approximations of the proposed tests, with approaches to minimize test statistic variance and to tune hyperparameters for maximum power. 

Our contributions are three-fold: 
\begin{enumerate}
    \item We formally describe tests on set-valued data, and to the best of our knowledge for the first time.
    \item We demonstrate the consistency of these tests for the two-sample and independence testing problems.
    \item We validate the proposed tests and optimization routines on simulated experiments that show that one may consistently discriminate between hypotheses on data that was previously not amenable to hypothesis testing.
\end{enumerate} 



\section{Background}
The tests presented in this paper are formally defined on distributions. Testing on distributions is the problem of defining a test statistic that maps distributions to a scalar that quantifies the evidence for a hypothesis we might set on the relationships in data. However, we do not have access to probability distributions themselves, but rather distributions are observed only through samples,
\begin{align}
    \{x_{1,j}\}_{j=1}^{n_1}, ... , \{x_{N,j}\}_{j=1}^{n_N}.
\end{align}
Each $\{x_{i,j}\}_{j=1}^{n_i}$ is a \textit{set} of $n_i$ individual observations $x_{i,j}$ (typically in $\mathbb R^d$). We assume that $\{x_{i,j}\}_{j=1}^{n_i}$ are $i.i.d$ samples from an unobserved probability distribution $\mathbb P_i$. The probability distributions $\{\mathbb P_i\}_{i=1}^N$ themselves have inherent variability, such as can be expected for example from different medical patients. We assume each one of them to be drawn randomly from some unknown meta-distribution $\mathcal M_P$ defined over a set of probability measures $\mathcal P$. We illustrate this set-up in Figure \ref{overview} for the two-sample problem (more details in Section \ref{two_sample_problem}).

\subsection{Embeddings of Distributions}
Let $\mathcal X$ be a measurable space of observations. We use a positive definite bounded and measurable kernel $k : \mathcal X \times \mathcal X \rightarrow \mathbb R$ to represent distributions $\mathbb P_i$ on $\mathcal X$, and independent samples $\{x_{i,j}\}_{j=1}^{n_i}$, as two functions $\mu_{\mathbb P_i}$, and $\hat\mu_{\mathbb P_i}$ respectively, called kernel mean embeddings \cite{muandet2016kernel}. Both are defined in the corresponding Reproducing Kernel Hilbert Space (RKHS) $\mathcal H_k$ by,
\begin{align*}
    \mu_{\mathbb P_i}:= \int_{\mathcal X} k(x,\cdot)d\mathbb P_i(x), \qquad \hat\mu_{\mathbb P_i}:= \frac{1}{n_i}\sum_{x \in \{x_{i,j}\}_{j=1}^{n_i}} k(x,\cdot)
\end{align*}
To make inference on populations of distributions, the desiratum however is on defining useful representations of distributions $\mathcal M_P$ on the space probability measures, rather than on the space of observations. \cite{christmann2010universal} showed that one may do so analogously to the definition of kernels on $\mathcal X$ by treating mean embeddings $\mu_{\mathbb P}$ themselves as inputs to kernel functions (replacing $x\in\mathcal X$ in the conventional learning setting as inputs to $k$). See eq. (\ref{eq3}) below. 

\textbf{Accounting for variance in embedding approximations.} In practice, each set representation $\mu_{\mathbb P_i}$ is limited to be approximated by irregularly sampled observations $\{x_{i,j}\}_{j=1}^{n_i}$. Not all mean embeddings $\mu_{\mathbb P}$ are expected to provide the same amount of information about their underlying distribution $\mathbb P$. Indeed, the empirical mean embeddings $\hat\mu_{\mathbb P_i}$ converge to their population counterpart at a rate $\mathcal O(1/\sqrt{n_i})$ (see e.g. Lemma 1 in the Appendix and also \cite{sriperumbudur2012empirical}) in their set size $n_i$. Rather than assuming access to a uniform sample of distributions $\{\mathbb P_i\}_{i=1}^N$ from $\mathcal M_P$, like we did with the raw observations $\{x_{i,j}\}_{j=1}^{n_i}$, we may account for this irregularity and uncertainty in approximation by interpreting the set of distributions as a weighted sample $\{(\mathbb P_i,w_i)\}_{i=1}^N\sim \mathcal M_P$. Each weight quantifying the accuracy of the approximation of each distribution with the limited samples available. The corresponding population and empirical mean embedding in this space may be written as,
\begin{align}
\label{eq3}
    \mu_{\mathcal M}:= \int_{\mathcal P} K(\mu_{\mathbb P},\cdot)d\mathcal M(\mathbb P), \qquad \hat\mu_{\mathcal M}:= \sum_{i=1}^N w_i K(\mu_{\mathbb P_i},\cdot)
\end{align}
We will make use of the Gaussian kernel between distributions defined $K(\mu_{\mathbb P},\mu_{\mathbb Q}):= \exp(-||\mu_{\mathbb P}-\mu_{\mathbb Q}||^2_{\mathcal H_K}/2\sigma^2)$ \cite{christmann2010universal,muandet2012learning}. Note that for kernels on $\mathcal X$, their RKHS consists of functions $\mathcal X \rightarrow \mathbb R$, while the kernel $K$ lives on the space of distributions on $\mathcal X$, $\mathcal{P(X)}$, and its RKHS consists of functions $\mathcal{P(X)}\rightarrow \mathbb R$. We may use $K$ to learn from samples that are individual distributions, rather than individual observations \cite{christmann2010universal}.

\textbf{Relationships with learning on distributions.} With this construction (i.e. kernels evaluated on mean embeddings) \cite{szabo2015two} investigated generalization performance in distributional regression: regressing to a real-valued response from a probability distribution. Results that were subsequently extended to study distributional regression for causal inference \cite{lopez2013randomized} and for transfer learning \cite{blanchard2017domain}. A technical contribution of this paper is to extend these results to demonstrate consistent hypothesis testing on distributions.

\subsection{Hypothesis Testing with Kernels}
The advantage for hypothesis testing of mapping distributions $\mathcal M$ and $\mathcal M'$ to functions in an RKHS is that we may now say that $\mathcal M$ and $\mathcal M'$ are close if the RKHS distance $||\mu_{\mathcal M}-\mu_{\mathcal M'}||_{\mathcal H_K}$ is small \cite{gretton2012kernel}. This distance depends on the choice of the kernel $K$ and $k$; a crucial property of the embeddings is that for certain kernels the feature map is injective. These kernels are called characteristic \cite{sriperumbudur2011universality}. Probability distributions may be distinguished exactly by their images in the RKHS, and also $||\mu_{\mathcal M}-\mu_{\mathcal M'}||_{\mathcal H_K}$ is zero if and only if the distributions coincide \cite{gretton2012kernel}. From the statistical testing point of view, this coincidence axiom is key as it ensures consistency of comparisons for any pair of different distributions. 

As a key property of the set-up we have introduced, in Theorem 2.2 \cite{christmann2010universal} demonstrated that for well known kernels, such as the Gaussian kernels, if used in both levels of the embedding and defined on a compact metric space the resulting embedding is injective (i.e. kernels are characteristic)\footnote{Theorem 2.2 \cite{christmann2010universal} technically shows that such kernels are universal, but universal kernels on compact metric space are known to be characteristic, as shown in Theorem 1 \cite{gretton2012kernel}.}.

The empirical version of the RKHS distance, however, will not necessarily be exactly zero even if the distributions do coincide. Some variability is to be expected due to the limited number of samples, and in contrast to conventional kernel tests, in the case considered here also due to the variability in the estimation of set embeddings. Instead of testing on an $i.i.d.$ sample $\{\mu_{\mathbb P_i}\}^N_{i=1}$, we are testing over the set $\{\hat\mu_{\mathbb P_i}\}^N_{i=1}$. There is an additional level of uncertainty which must be accounted for. 

In practice, tests are constructed such that a certain hypothesis is rejected whenever a test statistic exceeds a certain threshold away from 0 \cite{lehmann2006testing}. Then, short from achieving perfect discrimination between two hypotheses, the goal of hypothesis testing is to derive a threshold such that false positives are upper bounded by a design parameter $\alpha$ and false negatives are as low as possible. 


\subsection{Related work}
\textbf{Distances on sets}. As a first observation, note that kernels defined on sets directly \cite{kondor2003kernel}, measuring the similarity between sets by the average pairwise point similarities between the sets, are not known to be characteristic. Attempts have also been made to define kernels on the space of distributions, including probability product kernel \cite{jebara2004probability}, the Fisher kernel \cite{jaakkola1999exploiting}, diffusion kernels \cite{lafferty2005diffusion} and kernels arising from Kullback-Leibler divergences \cite{moreno2004kullback}, none of them known to be characteristic and in this case with the shortcoming that many of the above are parametrized by a family of densities which may or may not hold in data.

\textbf{Possible extensions to other tests}. Deep learning has emerged as an alternative for defining tests on structured objects. \cite{lopez2016revisiting} define classifier two-sample tests and \cite{liu2020learning} use deep kernels to embed structured objects. Tests in these cases, however, are defined directly on the space of observations, it is not clear how to input examples of varying sizes, or how to account for the uncertainty in individual observations especially if these change across sets. 

\textbf{Other conenctions with hypothesis testing.} Accommodating for input uncertainty has connections with robust hypothesis testing. These tests attempt to explicitly enforce invariances in test statistics in a certain uncertainty ball to remove irrelevant sources of variation \cite{gao2018robust,gul2016robust}. Other types of invariances can also be enforced, for instance \cite{law2017testing} use features designed to be invariant to additive noise and use distances between those representations for hypothesis testing. One may also use a model-based approach to capture this uncertainty, for instance \cite{benavoli2015gaussian} use Gaussian processes and compare posterior distributions. More generally, also work in the functional data analysis literature \cite{zhang2011statistical,panaretos2010second} uses a model-based approach to testing sets that represent functions.

\section{Hypothesis Tests on Sets}
In the following sections, we propose tests to evaluate two common hypotheses: the two sample problem of testing equality of distributions in two samples, and the independence problem of testing whether joint distributions in paired samples coincide with the product of their marginals. 

For both tests, the exposition mirrors well-known results in kernel hypothesis testing which we will only briefly describe (see \cite{gretton2012kernel,gretton2008kernel} for more background). The contribution of this paper is to show that tests defined with a second level of sampling are consistent and to show that correctly weighting representations according to their set size is most efficient.

\textbf{Algorithm.} We may summarize hypothesis testing in this context as follows:
\begin{enumerate}
    \item Embed the distributions $\{\mathbb P_i\}_{i=1}^N$ into an RKHS using approximations of the mean embeddings $\{\hat \mu_{\mathbb P_{i}}\}_{i=1}^N$ computed with independent samples $\{x_{i,j}\}_{j=1}^{n_i}\sim \mathbb P_i$.
    \item Define test statistics on this feature representations to test for a certain hypothesis or dependency structure in $\mathcal M$.
\end{enumerate}

\subsection{The two sample problem}
\label{two_sample_problem}
Consider a first collection of sets of observations, each $i$-th set denoted $\{x_{i,s}\}_{s=1}^{n_i}\sim \mathbb P_i$, for a total of $N$ such sets with distributions $\{\mathbb P_i\}_{i=1}^N \sim \mathcal M_P$, and define similarly a second collection of sets, each $j$-th set $\{y_{j,s}\}_{s=1}^{n_j}\sim \mathbb Q_j$, for $\{\mathbb Q_j\}_{j=1}^M \sim \mathcal M_Q$. The problem we consider is to test whether,
\begin{align}
    \mathcal H_0: \mathcal M_P = \mathcal M_Q \quad\text{or else}\quad \mathcal H_1: \mathcal M_P \neq \mathcal M_Q
\end{align}
holds on the basis of the observations available in each set. We illustrate this problem in Figure \ref{overview}. The proposed test statistic approximates the square of the RKHS distance between densities $\mathcal M_P$ and $\mathcal M_Q$, also called Maximum Mean Discrepancy (MMD), which may be decomposed as follows \cite{gretton2012kernel},
\begin{align}
    \MMD^2 :=  
    \mathbb E_{\mathbb P,\mathbb P' \sim\mathcal M_P}K(\mathbb P,\mathbb P') + \mathbb E_{\mathbb Q,\mathbb Q' \sim\mathcal M_Q}K(\mathbb Q,\mathbb Q') - 2 \mathbb E_{\mathbb P \sim\mathcal M_P, \mathbb Q \sim \mathcal M_Q}K(\mathbb P,\mathbb Q)
\end{align}
where $K$ is the kernel on distributions given after equation (\ref{eq3}). We denote $\widehat{\MMD}^2$ the empirical estimator of the $\MMD^2$ with expectations replaced by averages, obtained from independent samples $\{\mathbb P_i\}_{i=1}^N \sim \mathcal M_P$ and $\{\mathbb Q_j\}_{j=1}^M \sim \mathcal M_Q$. The proposed statistic is defined by considering approximate mean embeddings of each distribution and considering the weighted sample of their meta-distribution each of them represents,
\begin{align*}
    \widehat{\RMMD}^2 :=\sum_{i,j=1}^N  w_{\mathbb P_i}w_{\mathbb P_j}K(\hat\mu_{\mathbb P_i},\hat\mu_{\mathbb P_j}) + \sum_{i,j=1}^M w_{\mathbb Q_i}w_{\mathbb Q_j}K( \hat\mu_{\mathbb Q_i}, \hat\mu_{\mathbb Q_j})  - 2 \sum_{i,j=1}^{N,M} w_{\mathbb P_i}w_{\mathbb Q_j}K(\hat\mu_{\mathbb P_i},\hat\mu_{\mathbb Q_j})
\end{align*}
R stands for robust. Assume for now that all weights are fixed $w_{\mathbb P_i} = 1/N, w_{\mathbb Q_j} = 1/M$ for all $i,j$. We return to the specification of weights in section \ref{sec_practical}. The asymptotic behaviour of $\widehat{\MMD}^2$ is well understood \cite{gretton2012kernel} and the test itself is extensively used in many applications \cite{lloyd2015statistical,raj2019differentially}. However, these results do not extend trivially if each independent set exhibits an additional source of variation due to the estimation of the mean embedding. In the following proposition, we bound the contribution of this additional source of variation and show that under the asymptotic regime where both the set sizes and number of sets grow larger, asymptotic distributions are well defined.

\textbf{Proposition 1} \textit{\textbf{(Asymptotic distribution)}. Let two samples of data be defined as above and let $K$ be characteristic and $L_K$-Lipschitz continuous. Then, under the null and alternative and in the regime of increasing set size $n_i$ and increasing sample size $n$, the asymptotic distributions of $\widehat{\RMMD}^2$ coincides with that of $\widehat{\MMD}^2$.}  


\textit{Proof.} All proofs are given in the Appendix.
 
In other words, the additional variability due to a second level of sampling converges to 0 asymptotically, and thus the asymptotic distribution coverges to that of the well known MMD two sample test of \cite{gretton2012kernel}.

\subsection{The independence problem}
Independence tests are concerned with the question of whether two random variables are distributed independently of each other. For this problem, we start with a collection of \textit{paired} distributions $\{(\mathbb P_i,\mathbb Q_i)\}^N_{i=1}$ drawn from a joint distribution we denote $\mathcal M_{PQ}$, and denote their marginals $\mathcal M_{P}$ and $\mathcal M_{Q}$. The hypothesis problem is to determine whether,
\begin{align}
\label{independence_problem}
    \mathcal H_0: \mathcal M_{PQ} = \mathcal M_{P}\mathcal M_{Q} \quad\text{or else} \quad\mathcal H_1: \mathcal M_{PQ} \neq \mathcal M_{P}\mathcal M_{Q}
\end{align}

\textbf{Example.} Consider an example from healthcare to illustrate this problem.
\begin{itemize}[leftmargin=*]
    \item  A similar set-up as that given in Figure \ref{overview} may be used to illustrate independence testing with set-valued data. A common problem is identify dependencies between biomarkers, often observed irregularly over time in many patients. For instance cholesterol levels $\{x_{i,t_1}, \dots, x_{i,t_{n_i}}\}$ and blood pressure $\{y_{i,t_1}, \dots, y_{i,t_{n_i}}\}$ may be observed over times $t_1, \dots, t_{n_i}$ in $N$ individuals $i=1,\dots,N$. To formally test for dependencies between these samples one must account for the irregularity in observation time and uncertainty in biomarker reads. This can be done by considering instead distributions $\mathbb P_i$ and $\mathbb Q_i$ and testing for independence in this space directly.
\end{itemize}


As in the two-sample test, we may quantify the difference between distributions using the RKHS distance $||\mu_{\mathcal M_{PQ} } - \mu_{\mathcal M_{P} }\otimes \mu_{\mathcal M_{Q} }||^2_{HS}$. Kernels $K$, $L$ are assumed characteristic; $||\cdot ||_{HS}$ is the norm on the space of $\mathcal H_K \rightarrow \mathcal H_L$ Hilbert-Schmidt operators, and $\otimes$ denotes the tensor product, such that $(a \otimes b) c = a \langle b, c \rangle$. This distance is called the Hilbert Schmidt Independence Criterion (HSIC) \cite{gretton2005measuring,gretton2008kernel}.

Two empirical estimators can be written: one assuming access to independent samples $\mathcal M_{PQ}$ and one with independent samples from each of the paired distributions sampled from $\mathcal M_{PQ}$,
\begin{align}
    \widehat{\HSIC} =\Tr(KHLH)/N^2 \qquad\widehat{\RHSIC} =\Tr(\hat KH\hat LH)\cdot N^2
\end{align} 
for kernel matrices with $(i,j)$ entries $K_{ij}=K(\mathbb P_i,\mathbb P_j)=\langle\mu_{\mathbb P_i},\mu_{\mathbb P_j} \rangle_{\mathcal H_K}$ and $L_{ij}=\langle\mu_{\mathbb Q_i},\mu_{\mathbb Q_j} \rangle_{\mathcal H_L}$ for the population version and $\hat K_{ij}=w_{\mathbb P_i}w_{\mathbb P_j}\langle\hat\mu_{\mathbb P_i},\hat\mu_{\mathbb P_j} \rangle_{\mathcal H_K}$ and $\hat L_{ij}=w_{\mathbb Q_i}w_{\mathbb Q_j}\langle\hat\mu_{\mathbb Q_i},\hat\mu_{\mathbb Q_j} \rangle_{\mathcal H_L}$ with mean embeddings replaced by their weighted finite sample counterparts for the robust alternative. Assume for now that all weights are fixed $w_{\mathbb P_i} = 1/N, w_{\mathbb Q_j} = 1/M$ for all $i,j$. The centering matrix is defined by $H = I - \frac{1}{N} \mathbf{11}^T$ and Tr is the trace operator.

Here, similarly to the two sample problem, approximations due to a second level of sampling are well behaved and mirror those of the robust statistic for the two-sample problem. In particular, that asymptotic distributions of the RHSIC and the HSIC coincide in the regime with increasing set size and increasing sample size, making hypothesis testing with the $\widehat{\RHSIC}$ consistent for the independence problem in equation (\ref{independence_problem}). 

\textbf{Proposition 2} \textit{\textbf{(Asymptotic distribution)}. Let two samples of data be defined as above and let $K$ be characteristic and $L_K$-Lipschitz continuous. Then, under the null and alternative and in the regime of increasing set size $n_i$ and increasing sample size $n$, the asymptotic distributions of $\widehat{\RHSIC}$ coincides with that of $\widehat{\HSIC}$.}  

Independence testing with the $\widehat{\HSIC}$ has been studied in \cite{gretton2008kernel,zhang2018large,jitkrittum2017adaptive}.

\subsection{Practical remarks} 
\label{sec_practical}
We make a number of remarks on the practical application of our tests.

\begin{itemize}[leftmargin=*]
    \item \textbf{Weights for high power.} Set sizes in practice may be limited. In the asymptotic regime of increasing number of sets but finite set size, the properties of the estimator may depend on appropriately weighting sets for high power. The proposed weighting scheme addresses this point. 
    
    Recall that each individual observation $x_{ij}$ is drawn independently from their respective distributions $\mathbb P_i$. Other factors of variations assumed to be common across sets, the variance of the approximate embedding $\hat\mu_{\mathbb P_i}$ is therefore proportional to $1/n_i$ (i.e. the variation in approximation of mean embeddings is due solely to diverging set sizes). When mean embeddings have different variances, it is efficient to give less weight to mean embeddings that have high variances. By efficient in this context, we mean highest asymptotic power of tests based on mean embedding representations of sets. 
    
    For $V$-statistics the asymptotic power function is well known, and an argument involving the delta method for differentiable kernels, expanded on in the Appendix, can be used to determine the optimal weights to be given by $w_{\mathbb P_i} := n_i / \sum_i n_i$ for each $i$. 
    \item \textbf{Hyperparameters for high power.} With a similar intuition, even though in theory we can expect high power for any alternative hypothesis and any choice of kernel, with finite sample size, some kernel hyperparameters will give higher power than others. The proposed tests optimize the choice of kernels by choosing hyperparameters that minimize the asymptotic variance under the alternative similarly to \cite{sutherland2016generative,jitkrittum2017adaptive}. But, in addition, we extend the optimization to tune both the mean embedding to represent sets and the kernel used for comparisons in Hilbert space. Please find more details in the Appendix.
    \item \textbf{Low-dimensional approximations for large scale data.} Testing on distributions as described is often not scalable for even to large datasets, as computing each of the entries of the relevant kernel matrices requires defining a high-dimensional mean embedding. To define test statistics on these representations we further embed the non-linear feature space $\mathcal H_k$ defined by $k$ into a random low dimensional Euclidean space using their expansion in Hilbert space as a linear combination of the Fourier basis \cite{rudin1962fourier,rahimi2008random}. If we draw $m$ samples from the Gaussian spectral measure, we can approximate the Gaussian kernel $k$ by,
    \begin{align*}
        k(x,y)&\approx \frac{2}{m}\sum_{j=1}^m \cos(\langle \omega_j,x\rangle + b_j) \cos(\langle \omega_j,y\rangle + b_j) \\
        &= \langle\phi(x), \phi(y)\rangle 
    \end{align*}
    where $\omega_1,...,\omega_m \sim \mathcal N(0,\gamma)$, $b_1,...,b_m \sim \mathcal U[0, 2\pi]$, and $\phi(x)=\sqrt{\frac{2}{m}}[\cos(\omega_1 x + b_1),...,\cos(\omega_m x + b_m)]\in\mathbb R^m$ \cite{rahimi2008random}. The mean embedding $\mu_{\mathbb P}= \mathbb E_{X\sim \mathbb P} \phi(X)$ can then be approximated with elements in the span of $(\cos(\langle \omega_j,x\rangle + b_j))_{j=1}^m$. By averaging over the available $n_i$ samples in $X_i$ from the distribution $\mathbb P_i$, the approximate finite-dimensional embedding is given by,
    \begin{align*}
        \hat\mu_{\mathbb P_i,m} = \frac{1}{n_i}\sum_{x \in \{x_{ij}\}_{j=1}^{n_i}}\sqrt{\frac{2}{m}}(\cos(\langle w_j,x\rangle + b_j))^m_{j=1} \in \mathbb R^m
    \end{align*}
\end{itemize}


\section{Synthetic Data Experiments}
\label{exp}
The purpose of synthetic experiments will be to test \textbf{power}: the rate at which we correctly reject $\mathcal H_0$ when it is false, as we increase the difficulty of the testing problems; and \textbf{Type I error}: the rate at which we incorrectly reject $\mathcal H_0$ when it is true. 

In all experiments, $\alpha$ (the target Type I error) is set to $0.05$, the number of time series is set to $N=500$, the number of observations made on each time series is random between 5 and 50, and each problem is repeated for 500 trials. 

\textbf{Tests for empirical comparisons.} To the best of our knowledge, no existing test naturally accommodates for set-valued data with irregular sizes. Our approach to empirical comparisons will be to coerce the data into a fixed dimensional vector in a well-defined manner, and evaluate existing tests on this representation. To do so, we focus on time-series -like data which we interpolate along the time axis with cubic splines and evaluate at a fixed number of time points. 
\begin{itemize}[leftmargin=*]
    \item The following tests are evaluated for the two-sample problem. The \textbf{MMD} \cite{gretton2012kernel} with hyperparameters optimized for maximum power, two-sample classifier tests \cite{lopez2016revisiting} which involve fitting a deep classifier. We considered a recurrent neural network with GRU cells for sequential data (\textbf{C2ST-GRU}) and the DeepSets approach of \cite{zaheer2017deep} modelling permutation invariance to be expected in sets (\textbf{C2ST-Sets}). We consider also the Gaussian process-based test (\textbf{GP2ST}) by  \cite{benavoli2015gaussian}. 
    \item For the independence problem we consider: the \textbf{HSIC} \cite{gretton2008kernel}, the Randomized Dependence Coefficient (\textbf{RDC}) \cite{lopez2013randomized} and Pearson Correlation Coefficient (\textbf{PCC}). 
\end{itemize}
For all kernel-based tests, because their null distributions are given by an infinite sum of weighted $\chi^2$ variables (no closed-form quantiles), in each trial we use 400 random permutations to approximate the null distribution. We give more details on the implementation of each of these tests in the Appendix. 

\begin{figure*}[t]
\captionsetup{format=myformat}
\centering
\includegraphics[width=1\textwidth]{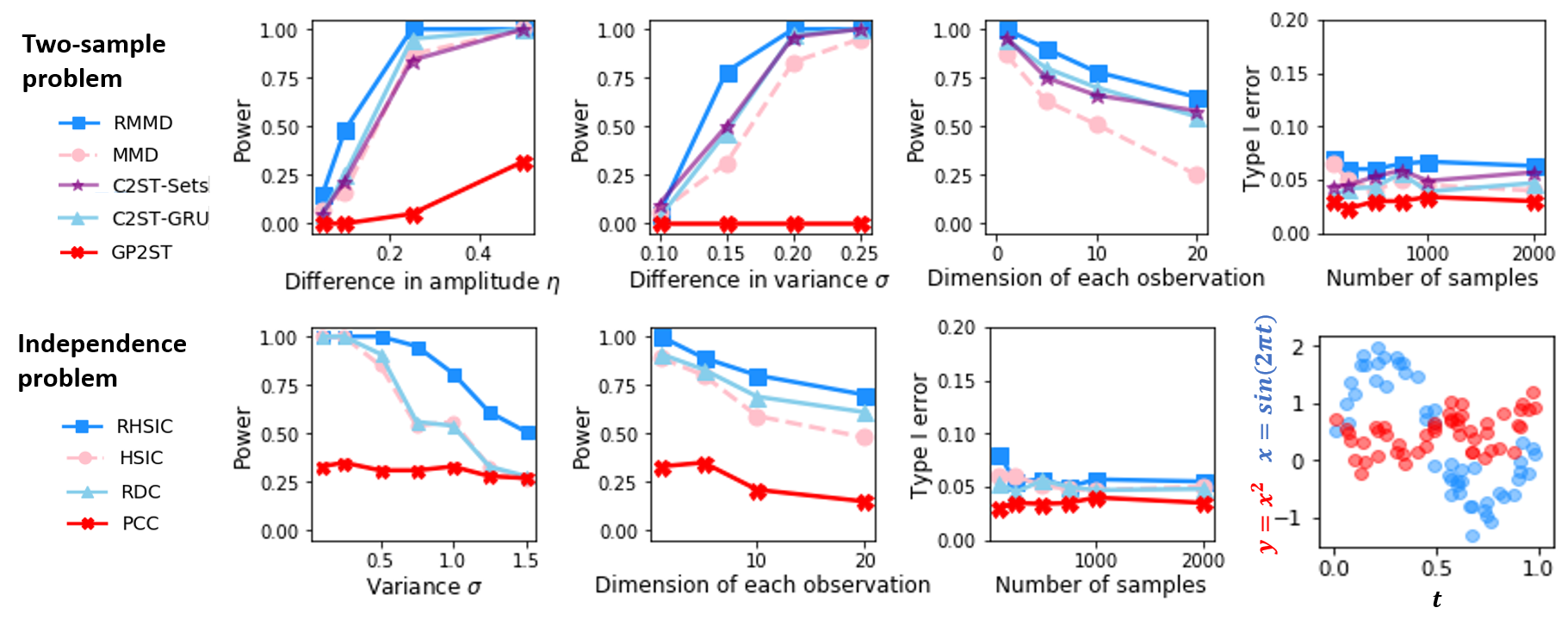}
\caption{Power (higher better) and Type I error on synthetic data. \textbf{Top row:} The top panels, from left to right, evaluate power as we increase the difference in time series amplitude (with equal variance $\sigma = 0.1$) and observation variance (with equal amplitude $\eta = 1$) between the two populations. Power comparisons as the dimension of each time series increases (on data sampled with a difference in amplitude equal to $0.25$) is shown next. Each new dimension is sampled as in the one-dimensional problem but with equal amplitude across the two populations, in other words only the distribution of the first dimension in each multivariate time series varies. The rightmost panel gives type I error with approximate control at the level $\alpha=0.05$ for all methods. \textbf{Bottom row}: The bottom row considers the independence problem, we evaluate power as we increase the variance of paired time series, and consider increasing dimensionality for a fixed variance $\sigma = 0.5$. Finally, the bottom right plot shows a sample of two dependent noisy time series, colored blue and red respectively, for illustration.}
\label{syn_fig}
\end{figure*}

\subsection{Two-sample problem}

\textbf{Experiment design.} Each one of the two samples is defined by a family of $N$ distributions $\{\mathbb P_i\}_{i=1}^N$ we take to be Gaussian $\mathbb P_i = \eta\sin(2\pi t) + \mathcal N(0,\sigma_i + \sigma)$. The variability between the $\{\mathbb P_i\}_{i=1}^N$ is specified by $\sigma_i$, drawn from a one-parameter inverse gamma distribution, which mimics the behaviour of the meta-distribution and the observation pattern we may observe in heterogeneous data. The difference between two populations of sampled distributions is the mean amplitude $\eta$ and/or shifts in \textit{baseline} variance $\sigma$. 

Two-sample problems become harder whenever these parameters converge to the same value in the two samples and are easier when they diverge. The sampled Gaussian distributions themselves are not observable and, in turn, we have access to observations $x_{ij} \sim \mathbb P_i$. Each $x_{ij}$ is obtained by fixing $t$ to $t_j \sim\mathcal U[0,1]$ and subsequently sampling from the Gaussian. 

The result is two collections of noisy time series with non-linear dynamics. Each time series, or set of observations, is irregularly sampled with noise levels that vary between sets. 

\textbf{Results.} We report power and type I error for the two sample problems in the top row of Figure \ref{syn_fig}. All tests approximately control for type I error at the desired threshold. In terms of power, we observe the RMMD to outperform across all experiments with an important contrast on the difference in performance with the MMD. Even though using similar test statistics, the RMMD much more faithfully captures the irregularity and uncertainty of every individual set of observations. RMMD similarly outperforms C2ST-based tests, the strongest baselines, with up to a two-fold increase in power in some cases.

\subsection{Independence problem} 

\textbf{Experiment design.} Define the mean of each distribution $\mathbb P_i$ as $f_i(t) := \beta_{i}\sin(2\pi t) + \alpha_{i} t$. Differently than in the two-sample problem, the variability among the $\{\mathbb P_i\}$ appears in the amplitude and trend of the sine function, let these be $\beta_i \sim \mathcal U[0.5,1.5]$ and $\alpha_i \sim \mathcal U[-0.5,0.5]$. Once these parameters are sampled, paired distributions $(\mathbb P_i,\mathbb Q_i)$ are given by $\mathbb P_i = f_i(t) + \mathcal N(0,\sigma)$ and $\mathbb Q_i = g(f_i(t)) + \mathcal N(0,\sigma)$. Each observation from this pair is obtained as in the two sample problem by fixing a random $t$ and sampling from the resulting distribution. 

The difficulty of the problem is governed by two factors: $g$ and $\sigma$. $g$ determines the dependency between the two functions. In every trial, $g(x)$ is randomly chosen from the set of functions $\{x^2,x^3,\cos(x),\exp(-x)\}$. Testing for dependency is hard also for increasing variance $\sigma$ of observations, as this makes the dependent paired samples appear independent. A sample of dependent sets of data using this data generating mechanism is given in the lower rightmost panel of Figure \ref{syn_fig}.

\textbf{Results.} Power and type I error are shown in the bottom row of Figure \ref{syn_fig}. The conclusions for this problem mirror the two-sample testing experiments, with however a much larger increase in power over alternatives, all using less flexible data representations as none of them avoids interpolating between observations before testing independence which we hypothesize is one reason for their underperformance. This is consistent with the increasing variance experiment, in this case increasing variance worsens interpolation performance. 

\begin{figure*}[t]
\centering
\includegraphics[width=1\textwidth]{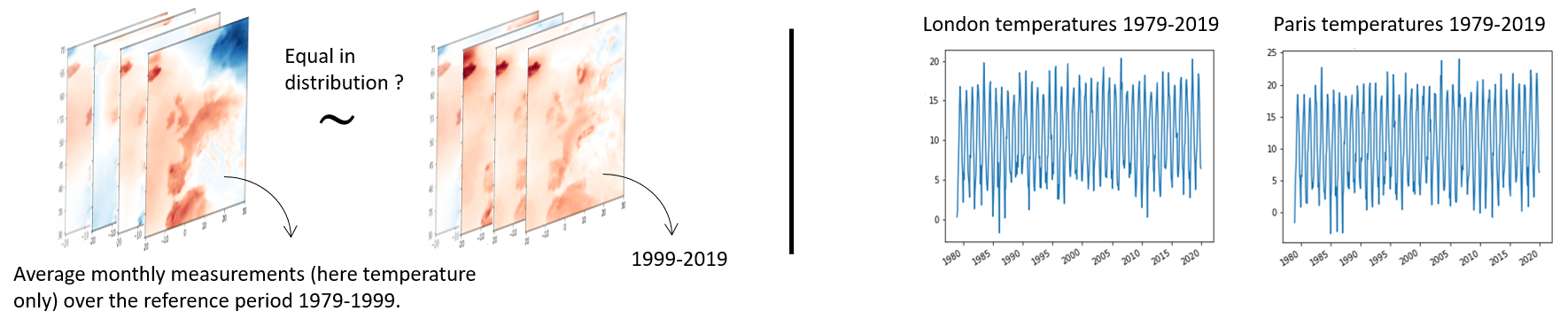}
\caption{Illustration of the two-sample problem with \textit{global} set-valued data versus \textit{local} time series data.}
\label{climate_data}
\end{figure*}

\section{Testing on Lung function Data of Cystic Fibrosis Patients}
For people with Cystic Fibrosis (CF), mucus in the lungs is linked with chronic infections that can cause permanent damage, making it harder to breathe \cite{kerem1992prediction}. This condition is often measured over time using \texttt{FEV1$\%$ predicted}; the Forced Expiratory Volume of air in the first second of a forced exhaled breath we would expect for a person without CF of the same age, gender, height, and ethnicity \cite{taylor2012understanding}. For example, a person with CF who has \texttt{FEV1$\%$ predicted} equal to 50$\%$ can breathe out half the amount of air as we would expect from a comparable person without CF. 
In this experiment, we work with data from the UK Cystic Fibrosis Trust containing records from $10,980$ patients with approximately annual follow ups between 2008 and 2015, with the objective of better understanding the dependence of lung function over time with other biomarkers. For this problem we found a significant influence of Body Mass Index (BMI) over time and the number of days under intravenous antibiotics in a given year; both already known to be associated with lung function \cite{wagener2018pulmonary,kerem2014factors}. 

\begin{figure}[H]
\centering
\includegraphics[width=0.45\textwidth]{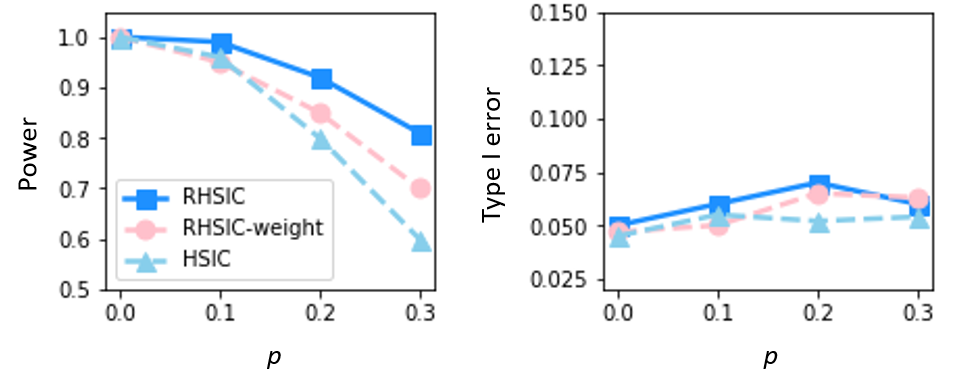}
\caption{Power and Type I error on Cystic Fibrosis data.}
\label{CF}
\end{figure}

We use this information to create a set of problems under the alternative $\mathcal H_1$ with an additional twist. We increase heterogeneity among patients by artificially removing a proportion $p$ of densely sampled patients (here more than 4 recordings). The problem is to test for independence between a patients two-dimensional trajectory of BMI and antibiotics measurements over time, and their lung function trajectory over time. In this set-up, we expect the information content of the average patient to decrease, a scenario that lends itself to an importance-weighted approach (more weight on densely sampled trajectories), such as described in section \ref{sec_practical}. In this section we test this property, which we found advantageous for higher missingness data patterns, as shown in Figure \ref{CF}. In this case, power tends to be higher after weighting (RHSIC) versus not weighting (RHSIC-weight). We report also type I errors, well controlled by all methods, evaluated after shuffling the lung function trajectories between patients, such as to break the associations between BMI and antibiotics, and lung function trajectories. 

\section{Testing on Climate Data}
This experiment explores the use of extensive weather data to determine whether the recent rapid changes in climate associated with human-induced activities significantly differ from natural climate variability. A number of variables are used to monitor the state of the climate including precipitation, wind patterns, and atmospheric composition among others. It depends on the latitude and longitude, and regions may vary and evolve differently over time. 

\textbf{Interpretation as set-valued data.} We can think of the multivariate measurements in different locations across the globe at a given time as a set of data points. Each set sampled from a probability distribution that represents the global weather pattern of the climate. We follow standard descriptions to define the climate as a collection of these sets observed over a period of 20 years. The problem is to test for significant differences in climate, represented by the evolution of bags of (multi-channel) images, over time (see Figure \ref{climate_data}).

\textbf{Experiment design.} The data is publicly available, provided by the Copernicus Climate Change Service\footnote{https://climate.copernicus.eu/.}. We include a total of 12 climate variables identified as essential to characterize the climate\footnote{https://public.wmo.int/en/programmes/global-climate-observing-system/essential-climate-variables}, including temperature, atmospheric pressure, observed over monthly periods for the last 40 years across Europe. The available data thus consists of a two streams of sets $\{x_{i,j}\}_{j=1}^{n_i}$ and $\{y_{i,j}\}_{j=1}^{n_i}$ for $i=1,\dots,144$ (12 months over 20 years). The first describes the climate over the period $1979-1999$, and the second set over the period $1999-2019$. Both contain measurements $x_{i,j}\in\mathbb R^{12}$ ($y_{i,j}$ respectively) in \textit{approximately} $n_i=250$ different locations (approximately because not all locations are consistently observed over time) which makes the length of each set irregular. Existing tests would thus require some form of interpolation which is not trivial over space and time in this case.

\textbf{Problem.} The problem is to test for the hypothesis of equally distributed climate data over the past 4 decades. We make different test: on data from the European, African, North American, South American and South-East Asian regions. 

\textbf{Results.} RMMD rejects the hypothesis of equally distributed climate data over the past 4 decades in Europe ($p$-value 0.0002), Africa ($p$-value 0.0014), and South America ($p$-value 0.0001) but fails to reject at a level of $0.01$ for North America ($p$-value 0.016) and South-East Asia ($p$-value 0.036). 

In the case of Europe, we note that this result would be different if only a particular location was considered (which could have been a viable reductionist strategy to use existing tests). For instance, we found that the RMMD applied to climate data over the same periods in London and Paris to not be significantly different ($p$-value 0.21). This experiment demonstrates the potential benefits of using more flexible tests that better represent available data to faithfully investigate complex phenomena such as climate that involve multiple measurements over time and space. 

\section{Conclusions}
In this paper we extended the toolkit of applied statisticians to do hypothesis testing on \textit{set}-valued data. We have shown that by appropriately representing each set of observations in a Hilbert space, kernel-based hypothesis testing may be applied consistently. Specifically, we introduced tests for the two-sample and the independence problem, derived their asymptotic distributions and provided efficient algorithms and optimization schemes to analyse a wide range of scenarios in an automatic fashion.

\bibliography{bibliography}
\bibliographystyle{plain}

\newpage
\appendix
{\Large \textbf{Appendix}}
\\\\
This appendix provides additional material accompanying the paper "Kernel Hypothesis Testing with Set-valued Data". It is outlined as follows:
\begin{itemize}
    \item Section \ref{sec_proofs} provides the proofs for all statements made in the main body of this paper.
    \begin{itemize}
        \item Section \ref{sec_proofs_twosample} gives the proof of the consistency of the RMMD.
        \item Section \ref{sec_proofs_independence} gives the proof of the consistency of the RHSIC.
    \end{itemize}
    \item Section \ref{sec_approximation} gives details on the approximations used to deal with irregular set sizes and high-dimensional data.
    \item Section \ref{sec_experiments} gives details on the synthetic experiments, implementation of baselines and source of the real data.
\end{itemize}

\section{Proofs}
\label{sec_proofs}

\subsection{Asymptotic distribution of $\widehat\RMMD^2$}
\label{sec_proofs_twosample}

Our proof strategy consists of demonstrating convergence in probability of each inner product $K(\hat\mu_{\mathbb P},\hat\mu_{\mathbb Q})$ to its population counterpart $K(\mu_{\mathbb P},\mu_{\mathbb Q})$, and take also into account approximations to the embeddings themselves we might make such as with Fourier features. Given convergence in probability, the equivalence of their asymptotic distributions then follows by convergence results of random variables. 

\subsubsection{Background}
All results in this section consider the asymptotic regime of increasing sample size $n$ and increasing set size $n_i$ for each $i$. We therefore make abstraction for notational purposes of our weighting mechanism, assumed fixed and each weight identical across sets asymptotically which is equivalent to reverting to the equal weight scenario for our asymptotic results.

We start by recalling some definitions. The empirical statistic of the RMMD is given by,
\begin{align}
    \widehat{\RMMD}^2 &:= \frac{1}{n^2}\sum_{i,j=1}^n K(\hat\mu_{\mathbb P_i}, \hat\mu_{\mathbb P_j}) + \frac{1}{m^2}\sum_{i,j=1}^m K( \hat\mu_{\mathbb Q_i}, \hat\mu_{\mathbb Q_j}) -  \frac{2}{nm}\sum_{i=1}^n\sum_{j=1}^m  K( \hat\mu_{\mathbb P_i}, \hat\mu_{\mathbb Q_j})
\end{align}
while the $\MMD$ with population mean embeddings is given by,
\begin{align}
    \widehat{\MMD}^2 := \frac{1}{n^2}\sum_{i,j=1}^n K(\mu_{\mathbb P_i}, \mu_{\mathbb P_j}) + \frac{1}{m^2}\sum_{i,j=1}^m K( \mu_{\mathbb Q_i}, \mu_{\mathbb Q_j}) -  \frac{2}{nm}\sum_{i=1}^n\sum_{j=1}^m  K( \mu_{\mathbb P_i}, \mu_{\mathbb Q_j})
\end{align}
We assume without loss of generality that $n=m$ for notational simplicity. 

Let us recall also the asymptotic distributions under the null and alternative of the $\widehat{\MMD}^2$ given by \cite{gretton2012kernel}.

\textbf{Theorem} \cite{gretton2012kernel}. Assume that $K$ has finite second moments. Then, the following statements hold. 
\begin{enumerate}[itemsep=0ex,topsep=0ex]
    \item Under $\mathcal H_0$, $n\widehat{\MMD}^2\hspace{0.2cm}\overset{d}{\operatorname{ \rightarrow }}\hspace{0.2cm} \sum_{l=1}^{\infty} \lambda_l(z^2_l - 2)$. $z_l$ is a sequence of gaussian random variables and $\lambda_l$ are the eigenvalues solution to a certain eigenvalue problem.
    \item Under $\mathcal H_1$, 
    $n^{1/2}\left(\widehat{\MMD}^2 - \MMD^2\right)\hspace{0.2cm} \overset{d}{\operatorname{ \rightarrow }}\hspace{0.2cm}\mathcal N \left(0,\sigma^2_{\mathcal H_1}\right)$.
\end{enumerate}
Please find the details of the eigenvalues and asymptotic variance in \cite{gretton2012kernel}.

Now note that,
\begin{align*}
    n\widehat{\RMMD}^2 &= n\widehat{\MMD}^2 + (n\widehat{\RMMD}^2 - n\widehat{\MMD}^2)\\
    \sqrt{n}\widehat{\RMMD}^2 &= \sqrt{n}\widehat{\MMD}^2 + (\sqrt{n}\widehat{\RMMD}^2 - \sqrt{n}\widehat{\MMD}^2)
\end{align*}
The first term relates to the asymptotic distribution of the RMMD under the null and the second term relates to the distribution of the RMMD under the alternative hypothesis.

We are interested in bounding the contribution of the second term in each case under the null and alternative hypotheses asymptotically. The absolute differences we are interested in bounding then under the null hypothesis given by,
\begin{align}
\label{null_bound}
    \left| n\widehat{\RMMD}^2 - n\widehat{\MMD}^2 \right| \leq & \frac{1}{n}\sum_{i,j=1}^n \left| K(\mu_{\mathbb P_i},\mu_{\mathbb P_j}) -K(\hat\mu_{\mathbb P_i},\hat\mu_{\mathbb P_j}) \right| \nonumber\\
    & + \frac{1}{n}\sum_{i,j=1}^n \left| K(\mu_{\mathbb Q_i},\mu_{\mathbb Q_j}) - K(\hat\mu_{\mathbb Q_i},\hat\mu_{\mathbb Q_j}) \right| \nonumber\\
    & - \frac{2 }{n}\sum_{i=1}^n\sum_{j=1}^n  \left| K(\mu_{\mathbb P_i},\mu_{\mathbb Q_j}) -K(\hat\mu_{\mathbb P_i},\hat\mu_{\mathbb Q_j}) \right|
\end{align}
and under the alternative hypothesis,
\begin{align}
\label{alternative_bound}
    \left|  \sqrt{n}\widehat{\RMMD}^2 -  \sqrt{n}\widehat{\MMD}^2 \right| \leq & \frac{ 1}{n\sqrt{n}}\sum_{i,j=1}^n \left| K(\mu_{\mathbb P_i},\mu_{\mathbb P_j}) -K(\hat\mu_{\mathbb P_i},\hat\mu_{\mathbb P_j}) \right|  \nonumber\\ 
    &+ \frac{ 1}{n\sqrt{n}}\sum_{i,j=1}^n \left| K(\mu_{\mathbb Q_i},\mu_{\mathbb Q_j}) - K(\hat\mu_{\mathbb Q_i},\hat\mu_{\mathbb Q_j}) \right| \nonumber\\
    & - \frac{2 }{n\sqrt{n}}\sum_{i=1}^n\sum_{j=1}^n  \left| K(\mu_{\mathbb P_i},\mu_{\mathbb Q_j}) -K(\hat\mu_{\mathbb P_i},\hat\mu_{\mathbb Q_j}) \right|
\end{align}
In both cases it suffices to show that inner products between population mean embeddings and empirical counterparts converge in probability at a rate fast enough such that a union bound over all terms in the summation scaled by $1/n$ and $1/(n\sqrt{n})$ converges to 0. We note here that we are considering two asymptotic regimes, once in the size of each set $n_i$ that is relevant in the convergence of $K(\hat\mu_{\mathbb P_i},\hat\mu_{\mathbb P_j})$ to $K(\mu_{\mathbb P_i},\mu_{\mathbb P_j})$ and one in $n$ which is the number of sets. Each may vary independently, and here we will assumed that the rate of growth of $n_i$ is sufficient to ensure the weighted sums converge as $n\rightarrow \infty$.

\subsubsection{Results}
We will traverse the convergence of empirical kernels to their population counterparts in two steps, first using results that show the convergence of empirical mean embeddings to their population counterparts (Lemma 1) and second, using a Lipschitz condition to extend this to inner products between mean embeddings (Lemma 2). 

For this we will assume $K$ to be a real-valued, shift invariant ($K(x,x')=K(x-x',0)$), and $L_K$-Lipschitz kernel,
\begin{align}
    |K(x,0) - K(x',0)| \leq L_K |x - x'|
\end{align}
also satisfying the boundedness condition $|K(x,x')|<1$ for all $x,x'\in\mathcal X$. 

The following two Lemmas demonstrate our claim.

\textbf{Lemma 1} (Bound on the empirical mean embedding \cite{lopez2015towards}) \textit{Let the kernel $K$ satisfy the assumptions above. Then we have,}
\begin{align}
    \left| \mu_{\mathbb P_i}-\hat\mu_{\mathbb P_i} \right |_{\mathcal H_K} \leq 2\sqrt{\frac{\mathbb E_{x\sim\mathbb P_i}K(x,x)}{n_i}} + \sqrt{\frac{2\log \frac{1}{\delta}}{n_i}}
\end{align}
\textit{with probability at least $1 - \delta$ over the randomness in the empirical sample from $\mathbb P_i$. $n_i$ is the number of samples from $\mathbb P_i$.}

\textbf{Lemma 2} (Bound on kernels computed on empirical mean embeddings) \textit{Let $K$ be defined as above. The it holds that,}
\begin{align}
    | K(\mu_{\mathbb P_i},\mu_{\mathbb P_j}) - K(\hat\mu_{\mathbb P_i},\hat\mu_{\mathbb P_j}) | \leq L_K \left( 4\sqrt{\frac{1}{\eta}} + 2\sqrt{\frac{2\log \frac{1}{\delta}}{\eta}} \right)
\end{align}
\textit{with probability at least $1 - \delta$. As $\eta:=\min(n_i,n_j) \rightarrow \infty$ we get that $K(\hat\mu_{\mathbb P_i},\hat\mu_{\mathbb P_j})$ converges in probability to $K(\mu_{\mathbb P_i},\mu_{\mathbb P_j})$.}

\textit{Proof.} The proof is based on the Lipschitz condition and the error bound on empirical mean embeddings with respect to their population counterparts.
\begin{align}
| K(\mu_{\mathbb P_i},\mu_{\mathbb P_j}) &- K(\hat\mu_{\mathbb P_i},\hat\mu_{\mathbb P_j}) |
= \left| K(\mu_{\mathbb P_i}-\mu_{\mathbb P_j},0) - K(\hat\mu_{\mathbb P_i}-\hat\mu_{\mathbb P_j},0) \right|\\
&\leq L_K \left| \mu_{\mathbb P_i}-\mu_{\mathbb P_j} - ( \hat\mu_{\mathbb P_i}-\hat\mu_{\mathbb P_j}) \right|\\
&\leq L_K \left| \mu_{\mathbb P_i}-\hat\mu_{\mathbb P_i} \right | + L_K\left | \mu_{\mathbb P_j}-\hat\mu_{\mathbb P_j} \right|\\
& \leq L_K \left( 2\sqrt{\frac{\mathbb E_{x\sim\mathbb P_i}K(x,x)}{n_i}} + \sqrt{\frac{2\log \frac{1}{\delta}}{n_i}} + 2\sqrt{\frac{\mathbb E_{x\sim\mathbb P_j}K(x,x)}{n_j}} + \sqrt{\frac{2\log \frac{1}{\delta}}{n_j}}  \right)\\
& \leq L_K \left( 4\sqrt{\frac{1}{\eta}} + 2\sqrt{\frac{2\log \frac{1}{\delta}}{\eta}} \right)
\end{align}
where $\eta := \min(n_i,n_j)$ and we have use the boundedness condition on $K$, $\mathbb E_{x\sim\mathbb P_i}K(x,x)\leq 1$. 

The for a rate of of increase of $n_i$ fast enough in comparison to $n$, each term in equations (\ref{null_bound}) and (\ref{alternative_bound}) converges to zero which implies that the asymptotic distributions of $n\widehat{\RMMD}^2$, $\sqrt{n}\widehat{\RMMD}^2$ and $\sqrt{n}\widehat{\MMD}^2$, $n\widehat{\MMD}^2$, coincide respectively.

\subsubsection{Extension to approximations using random Fourier features} 
For completeness, in addition to considering convergence in distribution using empirical embeddings, we extend our analysis to include Fourier feature approximations in the empirical embeddings themselves and their asymptotic behaviour. To do so notice that we may write,
\begin{align}
\label{decomp}
    | k(\mu_{\mathbb P_i},\mu_{\mathbb P_j}) - &k(\hat\mu_{\mathbb P_i,m},\hat\mu_{\mathbb P_j,m}) | \leq \nonumber\\
    &\left| k(\mu_{\mathbb P_i},\mu_{\mathbb P_j}) - k(\hat\mu_{\mathbb P_i},\hat\mu_{\mathbb P_j}) \right| + \left| k(\hat\mu_{\mathbb P_i},\hat\mu_{\mathbb P_j}) - k(\hat\mu_{\mathbb P_i,m},\hat\mu_{\mathbb P_j,m}) \right|
\end{align}
by the triangle inequality.

The following two lemmas are similar to the first two above but instead related the empirical mean embedding $\hat\mu_{\mathbb P_i}$ with its random Fourier feature approximation $\hat\mu_{\mathbb P_i,m}$.

\textbf{Lemma 3} (Bound on the randomized empirical mean embedding \cite{lopez2015towards}) \textit{Let $k$ be defined as above. For a fixed sample of size $n_i$ from a probability distribution $\mathbb P_i$ on $\mathbb R^d$ and any $\delta > 0$, we have,}
\begin{align}
    | \hat\mu_{\mathbb P_i} - \hat\mu_{\mathbb P_i,m} |_{L^2(\mathbb P)} \leq \frac{2}{\sqrt{m}}\left( 1 + \sqrt{2 \log n_i/\delta }\right)
\end{align}
\textit{with probability larger than $1-\delta$ over the randomness of the samples $(\omega_i,b_i)_{i=1}^m$.}

\textbf{Lemma 4} (Bound on kernels computed on approximated empirical mean embeddings) \textit{Let $k$ be defined as above. Then for any $\epsilon > 0$ it holds that,}
\begin{align}
    \left| k(\hat\mu_{\mathbb P_i},\hat\mu_{\mathbb P_j}) - k(\hat\mu_{\mathbb P_i,m},\hat\mu_{\mathbb P_j,m}) \right| \leq  \frac{2L_k}{\sqrt{m}}\left (2 + 2\sqrt{2\log(\eta/\delta)}\right)
\end{align}
\textit{$m$ is the number of random features, $n_i$ and $n_j$ are the number of observations in time series $X_i$ and $X_j$ respectively, and $\eta:=\min(n_i,n_j)$. If further we assume that $\min(n_i,n_j) \exp\{-m\} \rightarrow 0$ as $n_i,n_j,m \rightarrow \infty$, then $k(\hat\mu_{\mathbb P_i,m},\hat\mu_{\mathbb P_j,m})$ converges in probability to $k(\hat\mu_{\mathbb P_i},\hat\mu_{\mathbb P_j})$.}

\textit{Proof}. The proof strategy is similar to Lemma 3, but for with a different bound on the difference between mean embeddings. We proceed as follows,
\begin{align}
    \left| k(\hat\mu_{\mathbb P_i},\hat\mu_{\mathbb P_j}) - k(\hat\mu_{\mathbb P_i,m},\hat\mu_{\mathbb P_j,m}) \right| &= \left| k(\hat\mu_{\mathbb P_i}-\hat\mu_{\mathbb P_j},0) - k(\hat\mu_{\mathbb P_i,m}-\hat\mu_{\mathbb P_j,m},0) \right|\\
    &\leq L_k \left| \hat\mu_{\mathbb P_i}-\hat\mu_{\mathbb P_j} - ( \hat\mu_{\mathbb P_i,m}-\hat\mu_{\mathbb P_j,m}) \right|\\
    &\leq L_k \left| \hat\mu_{\mathbb P_i}-\hat\mu_{\mathbb P_i,m} \right | + L_k\left | \hat\mu_{\mathbb P_j}-\hat\mu_{\mathbb P_j,m} \right|\\
    &\leq \frac{2L_k}{\sqrt{m}}\left (2 + \sqrt{2\log(n_i/\delta)} + \sqrt{2\log(n_j/\delta)} \right)\\
     &\leq \frac{2L_k}{\sqrt{m}}\left (2 + 2\sqrt{2\log(\eta/\delta)}\right)
\end{align}
where we have written $\eta:=\min(n_i,n_j)$ and the inequalities hold with probability at least $(1-\delta)$ over the randomness of the samples $(\omega_i,b_i)_{i=1}^m$. 


\subsection{Asymptotic distribution of $\widehat\RHSIC$}
\label{sec_proofs_independence}

The asymptotic distribution of the RHSIC follows a very similar procedure since it can similarly bee decomposed in sums of kernels.

\textit{Proof.} The $\widehat\RHSIC$ may be written as a sum of $V$-statistics as follows \cite{gretton2008kernel},
\begin{align}
    \widehat\RHSIC = \frac{1}{N^2}\sum_{i,j}^N \hat K_{ij}\hat L_{ij} + \frac{1}{N^4}\sum_{i,j, q, r}^N \hat K_{ij}\hat L_{qr} - \frac{2}{N^3}\sum_{i,j, q}^N\hat K_{ij}\hat L_{iq}
\end{align}
where to avoid cluttering the notation we have written $\hat K_{ij}:= K(\hat\mu_{\mathbb P_i,m},\hat\mu_{\mathbb P_j,m})$ and $\hat L_{ij}:= L(\mu_{Y_i,m},\mu_{Y_j,m})$. Sums with two summation indices refer to double sums of all pairs of numbers drawn with replacement from $\{1,...,N\}$, and similarly for three and four summation indices \cite{gretton2008kernel}. Similarly to the two sample problem, equality in asymptotic distribution may be shown by considering the absolute differences in the product of population and empirical kernels. That is, we are interested in bounding the following,
\begin{align}
\label{diff_HSIC}
     |\hat K_{ij}\hat L_{qr} - K_{ij}L_{qr} | 
\end{align}
for any quadruple of indices $i,j,q,r$.

Assuming as above that kernels $K$ and $L$ are Lipschitz functions it follows that their product is also Lipschitz,
\begin{align*}
    | K(x,0)L(y,0) - &K(x',0)L(y',0) |  \\
    &\leq| (K(x,0)-K(x',0))L(y,0) + (L(y,0)-L(y',0))K(x',0)| \\ 
    &\leq
    |K(x,0)-K(x',0)| \cdot ||L(y,0)||_{\mathcal H_L} + |L(y,0)-L(y',0)| \cdot ||K(x',0)||_{\mathcal H_K}\\
    &\leq L_K |x - x'| + L_L |y - y'|
\end{align*}
The same arguments and lemmas used in the two-sample case apply which proves the equivalence in asymptotic distributions of the $\widehat\RHSIC$ and $\widehat\HSIC$.

\section{Approximations for high power}
\label{sec_approximation}

\subsection{Kernel hyperparameters}
For the two sample problem, let $N$ be the number of samples in both groups, which simplifies the formulation of the asymptotic power of the $\widehat{\RMMD}^2$. The following procedure mirrors \cite{sutherland2016generative}.

\textbf{Proposition 3} \textit{\textbf{(Approximate power of $\widehat{\RMMD}^2$)}. Under $\mathcal H_1$, for large $N$ and fixed $r$, the test power} $Pr(N\widehat{\RMMD}^2 > r)\approx 1 - \Phi(\frac{r}{\sqrt{N}\sigma_{\RMMD}} - \sqrt{N}\frac{\RMMD^2}{\sigma_{\RMMD}})$ \textit{where $\Phi$ denotes the cumulative distribution function of the standard normal distribution, $\sigma^2_{\RMMD}$ is the asymptotic variance under $\mathcal H_1$ for the $\widehat{\RMMD}^2$}.

Consider the terms inside the \textit{cdf} of the normal. Observe that the first term $\frac{r}{\sqrt{N}\sigma_{\RMMD}}= \mathcal O(N^{-1/2})$ goes to $0$ as $N \rightarrow\infty$, while the second term, $\sqrt{N}\frac{\RMMD^2}{\sigma_{\RMMD}} =  \mathcal O(N^{1/2})$, dominates the first one for large $N$. As an approximation, for sufficiently large $N$, the parameters that maximize the test power are given by $\theta^*= \text{argmax}_{\theta} \hspace{0.5em} Pr(N\widehat{\RMMD}^2 > r)\approx\frac{\RMMD^2}{\sigma_{\RMMD}}$. In our case $\theta$ includes the bandwidth parameter used to compute the mean embeddings and  the bandwidth parameter used to compute the test statistic. The empirical estimate of the variance $\hat \sigma_{\RMMD}$ that appears in our objective is approximated up to second order terms, as in \cite{sutherland2016generative}. Similar derivations hold for the power optimization of the HSIC with the exception that the definition of the HSIC requires optimization of two kernels, one for each set in our paired samples: $K$ and $L$.

Note that since RMMD and $\sigma_{\RMMD}$ are unknown, to maintain the validity of the hypothesis test we divide the sample into a training set, used to estimate the ratio with $\frac{\widehat\RMMD^2}{\hat\sigma_{\RMMD}}$ and choose the kernel parameters, and a testing set used to perform the final hypothesis test with the learned kernels.

An analogous result holds for the approximate power of $\widehat{\RHSIC}$.

\subsection{Weighting scheme}

Under the alternative hypothesis, the asymptotic variance of the proposed test statistics is well defined and given by asymptotic theory of $V$-Statistics (up to scaling) equal to $\Var(\mathbb E K(\mu_{\mathbb P_i},\mu_{\mathbb P_j}))$, see e.g. Theorem 5.5.1 \cite{serfling2009approximation}. To specify the set of weights that maximize power we may use the same reasoning to the section above and minimize the asymptotic variance. 

With finite samples to approximate the mean embedding, assuming that all randomness comes from the number of samples available to estimate mean embeddings, its variance is proportional to $1/n_i$. The delta method (see e.g. \cite{van1996delta}) may be applied on the bivariate sample $(\mu_{\mathbb P_i},\mu_{\mathbb P_j})$ with the function $K$ to conclude that the variance of each $K(\mu_{\mathbb P_i},\mu_{\mathbb P_j})$ is proportional to $1/(n_i\cdot n_j)$. Now, with a finite number of sets, or in other words a finite number of distributions, we approximate the expectation $\mathbb E K(\mu_{\mathbb P_i},\mu_{\mathbb P_j})$ with averages. Assuming that the covariance between any pair $K(\mu_{\mathbb P_i},\mu_{\mathbb P_j})$ and $K(\mu_{\mathbb P_i},\mu_{\mathbb P_k})$ for any $i,j,k$ does not vary by changing indices, that is, is fixed, weighting each term $K(\mu_{\mathbb P_i},\mu_{\mathbb P_j})$ with the inverse of its variance gives the lowest attainable variance $\Var(\mathbb E K(\mu_{\mathbb P_i},\mu_{\mathbb P_j}))$ in finite samples.


\section{Additional details on experiments and implementation}
\label{sec_experiments}

\subsection{Details on the data generation mechanisms}

The inverse gamma distribution has appeared parameterized by one and two parameters. We choose the one-parameter distributions with density,
\begin{align}
    f(x;\mu)=\frac{x^{-\mu-1}}{\Gamma(\mu)}\exp{(-1/x)}
\end{align}
where $x\geq 0$, $\mu > 0$ and $\Gamma$ is the gamma function.


\subsection{RMMD and RHSIC}

We create empirical kernel mean embeddings by concatenating data along each dimension. Each embedding has random features sampled to approximate a Gaussian kernel with length scale parameter $\sigma^2$. $\sigma^2$ is estimated by cross-validation on a grid of parameter values around the median of squared pairwise distances of the stacked data. In practice, we set the number of random features to $m = 50$ (larger amounts of random features show no significant performance improvements). The parameters of the kernel used for testing are similarly optimized via cross-validation by defining a grid of parameter values around the median of squared pairwise distances of computed random features. In summary, for each random feature length-scale we test with a number of test length-scales and choose the pair of parameters with best performance according to our power criterion. A summary of these tests' implementation is as follows.
\begin{enumerate}
    \item For each observed set $\{x_{i,j}\}_{j=1}^{n_i}\sim \mathbb P_i$, compute its approximated mean embedding using a Fourier basis, with elements in the span of $(\cos(\langle \omega_j,x\rangle + b_j))_{j=1}^m$,
    \begin{align*}
        \hat\mu_{\mathbb P_i,m} = \frac{1}{n_i}\sum_{x \in \{x_{ij}\}_{j=1}^{n_i}}(\cos(\langle w_j,x\rangle + b_j))^m_{j=1} \in \mathbb R^m
    \end{align*}
    \item Compute weights that describe the confidence we have in each of the above approximations, $w_{\mathbb P_i} := n_i / \sum_i n_i$ for each $i$, that result in posterior test statistics with lowest variance.
    \item Compute two-sample or independence test statistics on this weighted representation of the data to obtain a real-valued scalar $\hat t$ that discriminates between the two hypotheses of interest.
    \item In practice, a test decision will be made based on a comparison of the computed value $\hat t$ with an approximated null distribution obtained by repeated test statistic computation on permuted data representations. If $\hat t$ is greater than the $\alpha$ quantile of this approximated null distribution, reject the null hypothesis, otherwise fail to reject.
\end{enumerate}

\subsection{GP2ST}

The test developed by \cite{benavoli2015gaussian} was designed to test the equality of regression functions from observed two-dimensional data $(\mathbf t_{1}, \mathbf y_1)$ and $(\mathbf t_{2}, \mathbf y_2)$ from two samples. They assume a GP prior on the time series and compute posterior distributions by conditioning on each sample of observed data. Denote the posterior GPs by $f_1$ and $f_2$. With the assumption of gaussianity it follows that $\Delta f:= f_1 - f_2$ is also a GP, and evaluations on a fine grid of regular times $\mathbf t$ in $[0,1]$ will be multivariate Gaussian with mean denoted $\Delta\mu$ and covariance matrix $\Delta\Sigma$. The hypothesis of equality of data generating processes is then equivalent to testing departures of $\Delta f$ from the zero function. As a result, the two functions are equal with posterior probability $1 - \alpha$ if the credible region for $\Delta f$ includes the zero vector or, in other words, if:
\begin{align}
    \Delta\mu^T \Delta\Sigma^{-1}\Delta\mu \leq \chi^2_v(1-\alpha)
\end{align}
$\chi^2_v(1-\alpha)$ is the $(1-\alpha)$-quantile of a $\chi^2$ distribution with $v$ degrees of freedoms and $v$ is the number of positive eigenvalues of $\Delta\Sigma$.

\subsection{RDC}

The Randomized Dependence Coefficient (RDC) measures the dependence between fixed-dimensional random samples $X$ and $Y$ as the largest canonical correlation between $k$ randomly chosen nonlinear projections of their copula transformations. It is formally defined an analyzed in \cite{lopez2013randomized}.
   \begin{align*}
       \hat\rho (\mathbf x, \mathbf y) := \underset{\alpha,\beta}{\text{sup }} PCC(\alpha^T\Phi_{\mathbf x},\beta^T\Phi_{\mathbf y})
   \end{align*}
where $PCC$ is Pearson's correlation coefficient and $\Phi$ are nonlinear random projections, such as sine or cosine projections. To apply this function on irregularly observed data, we interpolate as we do with the MMD and HSIC.

We conduct a test using this measure of dependence by repeatedly shuffling the paired time series $M$ times to induce an empirical distribution of $\{\hat\rho_m\}_{m=1}^M$ under the null hypothesis of independence. The $p$-value is then given by $ \sum_{m=1}^M \mathbf{1}\{\hat\rho_m > \hat\rho\}/M$
where $\hat\rho$ is the statistic obtained from the observed data.

\subsection{PCC}
The Pearson's correlation coefficient (PCC) is a measure of linear correlation between two variables. It is defined as,
\begin{align*}
    \hat\rho (\mathbf x, \mathbf y) := \frac{\sum_i (x_i - \bar x)(y_i - \bar y)}{\sqrt{\sum_i(x_i - \bar x)}\sqrt{\sum_i(y_i - \bar y)}}
\end{align*}

Similarly to the RDC, we conduct a test using this measure of dependence by repeatedly shuffling the paired time series $M$ times to induce an empirical distribution of $\{\hat\rho_m\}_{m=1}^M$ under the null hypothesis of independence. 
   
\subsection{C2ST}

We implemented the C2ST with tensorflow in python. We used a RNN with GRU cells in one version and the deepset architecture (with the author's implementation \cite{zaheer2017deep}) in the other. The number of samples in each mini-batch is set to 64 the hidden layer size to 10. We optimize model parameters with Adam, learning rate equal to 0.01, and all variables are initialized with Xavier initialization. We use the elu activation functions for each layer and use sigmoid activation for the output layer given that we perform classification. 

Both tests proceeds as follows \cite{lopez2016revisiting}:

Let $\{x_i\}_{i=1}^n$ and $\{y_i\}_{i=1}^n$ be two samples of observed time series that include their corresponding time points in each case.
\begin{enumerate}
    \item Construct the data set $\mathcal D = \{(x_i,0)\}_{i=1}^n \cup \{(y_i,1)\}_{i=1}^n =: \{(z_i,l_i)\}_{i=1}^{2n}$.
    \item Shuffle $\mathcal D$ at random and partition into a training set $\mathcal D_{tr}$ and a testing set $\mathcal D_{te}$.
    \item Fit a classifier $g$ on the training set to predict the sample indicator $l$.
    \item Compute test statistic as classification accuracy on $\mathcal D_{te}$: $\widehat t := \frac{1}{n_{te}}\sum_{(z_i,l_i)\in \mathcal D_{te}}\mathbf 1\{\mathbf 1\{g(z_i)>1/2\}=l_i\}$
    \item If $\widehat t$ is greater that the $\alpha$ quantile of a $\mathcal N(1/2,1/(4n_{te}))$ reject $\mathcal H_0$; otherwise accept $\mathcal H_0$.
\end{enumerate}
$\mathbf 1$ is the indicator function.
\end{document}